\documentclass[
]{ceurart}

\usepackage{listings}
\usepackage{multirow}
\usepackage{cleveref}
\usepackage{amsthm}
\usepackage[utf8]{inputenc}
\usepackage{glossaries}
\usepackage{csquotes}

\usepackage[T1]{fontenc}
\usepackage[english]{babel}
\usepackage{microtype} 
\usepackage{tabularx} 
\usepackage{graphicx}
\usepackage{amsmath,amssymb,amsfonts}
\usepackage{stmaryrd}
\usepackage{textcomp}
\usepackage{xcolor}
\usepackage{mathtools}
\usepackage{pdfpages}

\usepackage{algorithm}
\usepackage{algpseudocode}

\usepackage{listings}
\newenvironment{ownfigure}[0]%
{\begin{figure}[htb!]}
{\end{figure}}

\usepackage{color}
\definecolor{DarkRed}{rgb}{0.75,0,0}
\definecolor{Lightgreen}{rgb}{0.588,1.0,0.588}
\definecolor{DarkGreen}{rgb}{0,0.5,0}
    
\lstdefinelanguage{MontiArc}[]{Java}{
  morekeywords={component, port, in, out, inv, package, import, connect, autoconnect}
}

\lstdefinelanguage{myJava}[]{Java}{
  commentstyle=\color{DarkGreen}\itshape 
}

\lstdefinelanguage{MontiArcAutomaton}[]{Java}{
  morekeywords={component, port, in, out, inv, package, import, connect,
  autoconnect, automaton, state, ocl, java, initial, final,
  noCompletion, chaosCompletion, var, mode, activate, transitions,
  modetransitions}, commentstyle=\color{DarkGreen}\itshape }

\lstdefinelanguage{MCConfig} { 
    morekeywords={config, Require, Model} 
}

\lstdefinelanguage{Manifest} { 
    morekeywords={Manifest, Bundle, ManifestVersion, Name, SymbolicName,
      Version, Require
    } 
}

\lstdefinelanguage{mcGrammar}[]{}{
  morekeywords={
    grammar, package, path, parser, lexer, nows, noslcomments, nomlcomments, 
    noident, nostring, noanything, nocharvocabulary, dotident, identrule,
    xmlcomments, hashcomments, texcomments, freemarkercomments, concept, 
    globalnaming, define, usage, options, true, false, protected, ident, 
    compilationunit
  }
}

\lstdefinelanguage{mcLng}[]{}{
  morekeywords={
    dsltool, language, package, path, parser, root, parsingworkflow, 
    rootfactory, lexer, nows, noslcomments, nomlcomments, noident, nostring,
    dotident, concept, globalnaming, define, usage, options, true, false, 
    protected, ident
  }
}

\lstdefinelanguage{mcManifest}[]{}{
  morekeywords={
    bundle, Bundle, Name, SymbolicName, true, false, Main, Class, 
    Version, Activator, Localization, Require, 
    Exclude, Eclipse, LazyStart, Vendor, Export, Package, 
    ClassPath
  }
}

\lstdefinelanguage{Alloy}[]{Java}{
commentstyle=\color{DarkGreen}\itshape,
  morekeywords={abstract,sig,->,fact,pred,fun,run,for,iff,
  not,no,one,all,some,lone,\#,set,in,and,or,but,exactly,none,univ,Int,assert,check},
  otherkeywords = {[2]????},
    morekeywords = {[2]????},
    keywordstyle={[2]\color{blue}},
    otherkeywords = {[3]????,<,<->,->, &, |, =, !=, !,<:,~},
    morekeywords = {[3]????,<,<->,->, &, |, =, !=, !,<:,~},
    keywordstyle={[3]\color{blue}}
  }

\lstdefinelanguage{mccd}[]{Java}{
  morekeywords={classdiagram,abstract,<<singleton>>,class,int,String,
  association,composition,extends}
}

\lstdefinelanguage{FreeMarker}[]{}{
  keywordsprefix={\#},
  keywords={in},
  commentstyle=\color{DarkGreen}\itshape }

\lstdefinelanguage{Mona}[]{}{
  morekeywords={ex0,all0,ex1,all1,ex2,all2,var0,var1,var2,pred,in,notin,include,union,inter,empty,assert},
  morecomment=[l]{\#},
  commentstyle=\color{DarkGreen}\itshape,
  otherkeywords = {[2]????,next,boolean,init,case,esac},
  morekeywords = {[2]????,next,boolean,init,case,esac},
  otherkeywords = {[3]????,<,<=>,=>, &, |, =, !=, !},
  morekeywords = {[3]????,<,<=>,=>, &, |, =, !=, !},
}

\lstdefinelanguage{myPython}[]{Python}{
  morekeywords={assert},
  morecomment=[l]{\#},
  commentstyle=\color{DarkGreen}\itshape,
}

\lstdefinelanguage{GeneratorConfiguration}[]{Java} {
  morekeywords={
    template, 
    generator, 
    ast, 
    runtime},
}

\lstdefinelanguage{ApplicationConfiguration}[]{Java} {
  morekeywords={
    application,
    behaviors,
    bindings,
    classdiagrams,
    components,
    factories,
    generators,
    map,
    to},
}

\lstdefinelanguage{Isabelle}[]{} {
    morekeywords={
        datatype,
        typedef},
}
\usepackage{xspace}
\usepackage{ifdraft}

\newcommand{\no}{\xmark\xspace}

\newcommand{\iow}{I/O$^\omega$\xspace}

\newcommand*{\eg}{\textit{e.g.,}\@\xspace}

\makeatletter
\newcommand*{\etc}{%
  \@ifnextchar{.}%
  {\textit{etc}}%
  {\textit{etc.}\@\xspace}%
}
\makeatother

\newcommand{\xmark}{\ding{55}}%
\definecolor{se-green}{RGB}{0,128,0}
\definecolor{se-blue} {RGB}{0,0,204}

\newcounter{requirement}[section]

\usepackage{amsthm}
\theoremstyle{definition}

\glsdisablehyper 

\newacronym{cc}{C\&C}{Component and Connector}

\newacronym{dsl}{DSL}{Domain Specific Language}

\newacronym{mda}{MDA}{Model-Driven Architecture}

\newacronym{cpu}{CPU}{Central Processing Unit}

\newacronym{FDDT}{FDDT}{formal description and development technique}

\newacronym{AFWME}{AFWME}{Apply Fluid with Mechanical Energy}

\newacronym{SRV}{SRV}{Set Rotational Velocity}

\newacronym{SD}{SD}{Synchronous Drive}

\newacronym{SI}{SI}{International System of Units (Système International d'unités)}

\newacronym{OMG}{OMG}{Object Management Group}

\newacronym{V\string&V}{V\string&V}{Validation and Verification}

\newacronym{TEE2ME}{TEE2ME}{Transform Electrical Energy to Mechanical Energy}

\newacronym{MDE}{MDE}{Model Driven Engineering}

\newacronym{ERS}{ERS}{Entity Relationship Schemata}

\newacronym{OCL}{OCL}{Object Constraint Language}

\newacronym{UML}{UML}{Unified Modeling Language}

\newacronym{UML/P}{UML/P}{\gls{UML}/P}

\newacronym{CD}{CD}{Class Diagram}

\newacronym{FD}{FD}{Feature Diagram}

\newacronym{OD}{OD}{Object Diagram}

\newacronym{OM}{OM}{Object Model}

\newacronym{AST}{AST}{Abstract Syntax Tree}
\def\ast{\gls{AST}\xspace}

\newacronym{SMT}{SMT}{Satisfiability Modulo Theories}

\newacronym{CPS}{CPS}{Cyber-Physical System}

\newacronym{cpf}{CPF}{Cyber-Physical Function}

\newacronym{ME}{ME}{Mechanical Engineering}

\newacronym{SE}{SE}{Software Engineering}

\newacronym{PDP}{PDP}{Product Development Process}

\def\sl{\textit{Simulink}\xspace}

\newacronym{OCL/P}{OCL/P}{OCL/Programmable}

\newacronym{sysml}{SysML}{Systems Modeling Language}

\newacronym{xml}{XML}{Extensible Markup Language}

\newacronym{ad}{AD}{Activity Diagram}

\newacronym{sc}{SC}{Statechart}
\def\sc{\gls{sc}\xspace}
\def\scs{\glspl{sc}\xspace}

\newacronym{ucd}{UCD}{Use Case Diagram}

\newacronym{uc}{UC}{Use Case}

\newacronym{sd}{SD}{Sequence Diagram}

\newacronym{bdd}{BDD}{Block Definition Diagram}

\newacronym{ibd}{IBD}{Internal Block Diagram}

\newacronym{mde}{MDE}{Model-Driven Engineering}

\newacronym{CAD}{CAD}{Computer-Aided Design}

\newacronym{SysML4FMArch}{SysML4FMArch}{SysML for Functional Mechanical 
Architectures}

\usepackage{ifdraft}
\usepackage{ifthen}

\newboolean{debug}
\setboolean{debug}{true}

\ifthenelse{\boolean{debug}}{%
	\usepackage{todonotes}                 
	\newcommand{\xynote}[2]{\todo[inline]{#1: #2}}
	\newcommand{\note}[1]{{\color{blue}\textit{#1}}}
	
	\usepackage{hyperref}                  
	\usepackage{setspace}                  

}{%
	\newcommand{\note}[1]{}
	\newcommand{\xynote}[2]{}
	
	\hypersetup{hidelinks}                 
}

\begin{document}


\copyrightyear{2022}
\copyrightclause{Copyright for this paper by its authors.
  Use permitted under Creative Commons License Attribution 4.0
  International (CC BY 4.0).}

\conference{arxiv.com}

\title{Dynamic Symbolic Execution for Semantic Difference Analysis of Component and Connector Architectures}

\author[1]{Johanna Grahl}[%
email=johanna.grahl@rwth-aachen.de
]
\fnmark[1]
\address[1]{RWTH Aachen University, Germany}

\author[2]{Bernhard Rumpe}[%
orcid=0000-0002-2147-1966,
email=rumpe@se.rwth-aachen.de,
url=https://se-rwth.de,
]
\fnmark[1]
\address[2]{Software Engineering, RWTH Aachen University, Germany}

\author[2]{Max Stachon}[%
orcid=0000-0002-6328-3816,
email=stachon@se-rwth.de,
]
\fnmark[1]

\author[2]{Sebastian Stüber}[%
orcid=0000-0002-6636-9375,
email=stueber@se-rwth.de,
]
\fnmark[1]

\fntext[1]{These authors contributed equally.}




\begin{abstract}
    In the context of model-driven development, ensuring the correctness and consistency of evolving models is paramount.
    This paper investigates the application of Dynamic Symbolic Execution (DSE) for semantic difference analysis of component-and-connector architectures, specifically utilizing MontiArc models. 
    We have enhanced the existing MontiArc-to-Java generator to gather both symbolic and concrete execution data at runtime, encompassing transition conditions, visited states, and internal variables of automata. 
    This data facilitates the identification of significant execution traces that provide critical insights into system behavior. 
    We evaluate various execution strategies based on the criteria of runtime efficiency, minimality, and completeness, establishing a framework for assessing the applicability of DSE in semantic difference analysis. 
    Our findings indicate that while DSE shows promise for analyzing component and connector architectures, scalability remains a primary limitation, suggesting further research is needed to enhance its practical utility in larger systems.
\end{abstract}


\maketitle

	\section{Introduction} \label{sec:introduction}
In model-driven software and systems engineering, models serve as the primary artifacts for development, evolving throughout their lifecycle due to modifications, refinements, and refactorings. To facilitate a less error-prone development process and ensure the preservation of critical model properties, automated model analyses, such as semantic differencing, can be employed \cite{MRR10}.

Semantic differencing is a comparative model analysis technique that evaluates two models written in the same language by examining their legal instances as defined by a language-specific formal semantics \cite{HR04}. A semantic difference is identified when a "\enquote{diff-witness} exists—an instance that is valid in one model but not in the other. Conversely, if no such witness is found, the first model can be considered a semantic refinement of the second. Various semantic difference operators have been developed for different modeling constructs, including activity diagrams \cite{MRR11d, KR18}, class diagrams \cite{MRR11b,RRS23}, feature models \cite{DKMR19}, variants of statecharts \cite{BKRW17,BKRW19,DEKR19}, and sequence diagrams \cite{Kau21}.

Despite the extensive exploration of semantic differencing in static structural models, such as class diagrams, and isolated behavioral models, like statecharts, component-and-connector architectures present unique challenges due to their dynamic nature and compositional complexity. Unlike static object structures that do not define dynamic behavior, component-and-connector models necessitate the consideration of both individual component behaviors and their interactions within compositions. This paper addresses these challenges by focusing on semantic differencing specifically for MontiArc models \cite{HRR12,Hab16,BKRW17a}, a topic that has yet to be comprehensively covered in existing literature.

Understanding changes in architectural models is essential for early bug detection and ensuring correct refinements from underspecified designs. Semantic differencing is particularly effective in addressing underspecification and refinement during the early stages of development, when architectural designs are abstract and not fully specified. As the design evolves, this initial underspecification is progressively refined through decomposition, architectural refactorings, and restrictions on component behavior. Our approach not only identifies semantic differences but also generates test cases that assist in subsequent development phases. These test cases provide valuable insights into specified behaviors and highlight areas that may require further specification or testing, thereby supporting a more robust model evolution process.

To analyze behavioral differences in component and connector architectures, we aim to leverage \textit{Dynamic Symbolic Execution (DSE)} \cite{Cadar}, a variant of symbolic execution \cite{king}, to develop a semantic differencing operator specifically for MontiArc models. Building upon the fundamental semantic-difference framework established in previous publications, we apply DSE as a novel technique for identifying semantic differences in these models.\label{review:semdiff_clarification}

MontiArc~\cite{HRR12,Hab16,BKRW17a} is an architecture modeling framework developed with the MontiCore\footnote{\url{https://monticore.github.io/monticore/}} language workbench~\cite{GKR+06,HKR21} that offers an architecture description language for designing both software and cyber-physical systems~\cite{GRSS12}.
It enables modular hierarchical modeling~\cite{BR07} of architectural components and their message-oriented, asynchronous communication. 
The input-output behavior of atomic components can then be specified by a variant of \scs~\cite{statechart,Rum16} based on the FOCUS formalism~\cite{BS12}.

To facilitate simulations, MontiArc includes a code generator that produces executable Java code \cite{Hab16,HRR12}. We have extended this generator to create symbolic code suitable for DSE. This symbolic code can be invoked using a newly developed tool designed to initiate the analysis. A default execution strategy for DSE is provided, with support for additional user-defined strategies. These strategies utilize execution data from the MontiArc model and employ Z3 \cite{efficientZ3} as a satisfiability modulo theories (SMT) solver.

The information collected during execution includes both symbolic and concrete values, which serve as inputs for execution strategies and contribute to model validation efforts. An exemplary use case for this model validation is the calculation of semantic differences between MontiArc models. Specifically, a semantic difference is identified by a \enquote{diff-witness}, which manifests as an input-output trace from the first model that cannot be replicated in the second model.

Our implementation supports primitive data types, strings, and enumeration types for ports and internal variables. Furthermore, it accommodates model composition, non-deterministic transition selection, and model parameters. Notably, we currently focus on component behavior specifications defined through \scs and time-synchronous communication between components.
    
\paragraph{Contributions} In this paper, we present the following contributions:

\begin{itemize}
\item Development and implementation of DSE for MontiArc.
\item Implementation of multiple DSE controllers, each employing distinct execution strategies.
\item Realization of a DSE-based semantic differencing operator specifically for MontiArc models.
\item Evaluation of the implemented controllers based on the criteria of \textit{runtime}, \textit{minimality}, and \textit{completeness}, as well as their applicability to semantic differencing.
\item Discussion on the utility and limitations of DSE in the context of component-and-connector architectures.
\end{itemize}

Our evaluation emphasizes completeness and runtime efficiency. However, it is important to note that enforcing completeness can lead to exponential increases in runtime with respect to input complexity. To address this limitation, we discuss potential mitigation strategies that can help balance these factors.

The remainder of this paper is structured as follows:
In Section \ref{sec:dseDefinition}, we outline the concept of DSE. This is followed by an exploration of related work in the field of DSE in Section \ref{sec:relatedWork}. Section \ref{sec:caseStudy} presents an example architecture to illustrate and evaluate our approach in the subsequent sections. The concept and design decisions behind our tool's development are detailed in Section \ref{sec:concept}.
In Section \ref{sec:evaluation}, we evaluate our DSE approach and the implemented execution strategies, followed by a discussion of the results and their implications in Section \ref{sec:discussion}. Finally, we conclude in Section \ref{sec:conclusion} with a summary of our findings and an outlook on future work.
    \section{Dynamic Symbolic Execution} \label{sec:dseDefinition}
Dynamic Symbolic Execution (DSE) is a variant of symbolic execution that generates concrete inputs alongside symbolic inputs~\cite{Cadar}. In symbolic execution, each input is represented as a symbolic constant. Each execution path of a program is associated with a boolean expression known as the path condition. These path conditions are initialized with the value \textit{true} and define the properties that must be satisfied by a concrete value in order to execute that specific path. During the process of symbolic execution, an \textit{execution tree} is constructed, representing all paths identified by the chosen execution strategy~\cite{king}. Concrete values can be derived as inputs by solving individual path constraints using a Satisfiability Modulo Theories (SMT) solver.

\begin{figure}[h]
\begin{lstlisting}[language = java, caption={Example of a program with an unsatisfiable path},label=example_program]
public int example(int x){
  int z = 2*x;
  if(z < 10){
    if(z > 10) { z = z + 1; } // unreachable
    z = z * 2;		
  }
  return z;	
}
\end{lstlisting}

		\begin{center}
			\includegraphics[width = 6cm , height = 6 cm, keepaspectratio]{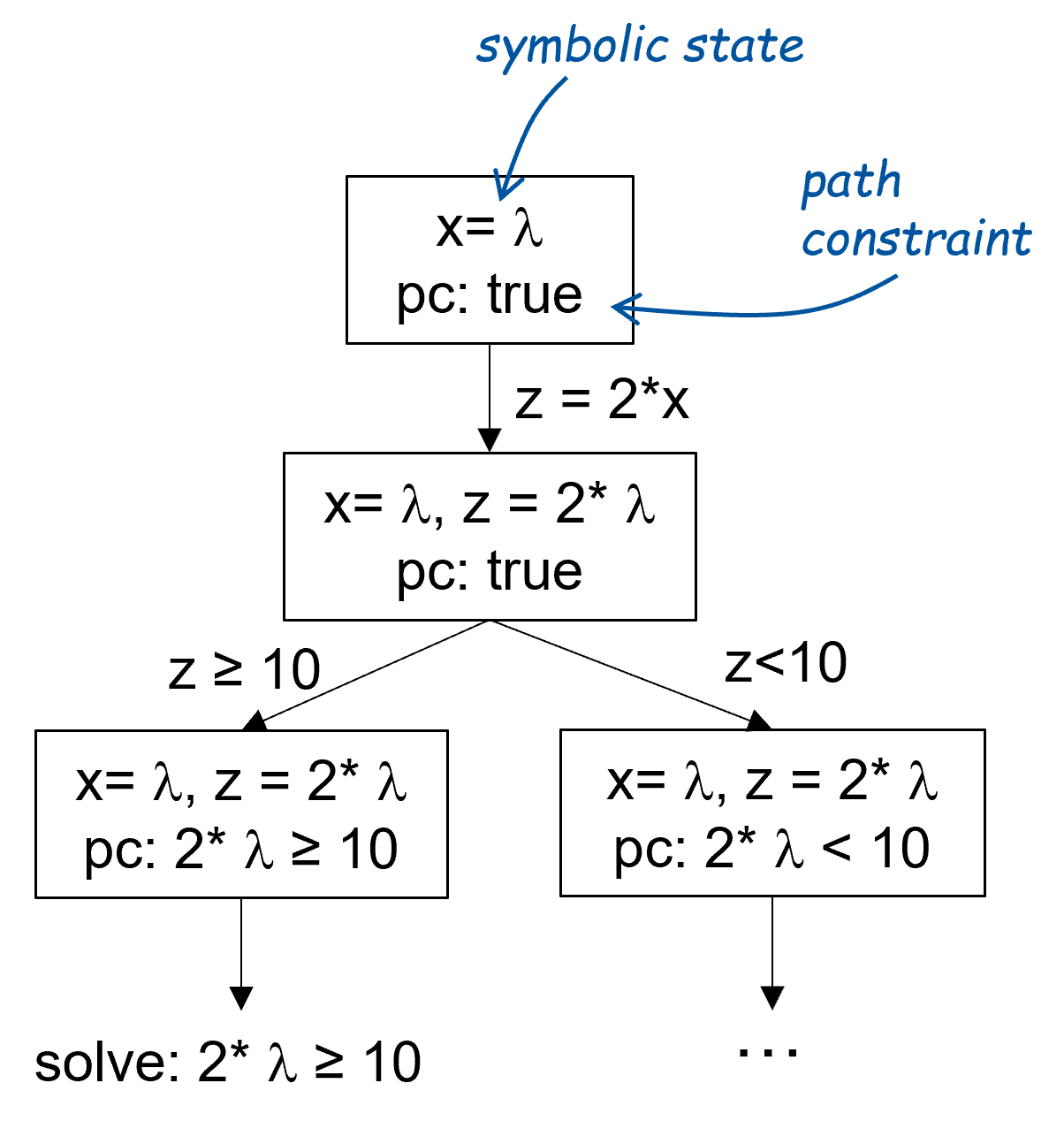}
			\caption{Symbolic execution tree. At the top is the initial state, arrows are assignments or branches.}
			\label{symbolic_execution_tree}
		\end{center}
	\end{figure}

Consider the program illustrated in \cref{example_program}, which takes an integer input \texttt{x}. In symbolic execution, an execution tree is constructed, and a symbolic constant (\eg $\lambda$) is applied to \texttt{x}. The path constraint is initialized with \textit{true}, as depicted in \cref{symbolic_execution_tree}. While executing the program, assignments are also executed symbolically; for example, the assignment $z=2*x$ (cf. line 2) would be represented as $z=\lambda*2$. At branching points, such as an \texttt{if} statement, the corresponding constraints are collected. The \texttt{if} condition in line 3 results in two possible paths, each with the following path constraints: $x<10$x or $x\geq10$. If the condition $x<10$ is not satisfied, the execution of that particular path in the execution tree will terminate. Conversely, the alternative path is analyzed to completion. Each leaf node of the execution tree corresponds to a path constraint, representing the specific condition that must be satisfied for that execution path to be taken.
	
Symbolic execution has inherent limitations, particularly in cases where certain functions cannot be effectively executed symbolically, such as cryptographic functions \cite{Godefroid_SMART}. Dynamic Symbolic Execution (DSE), also known as concolic execution, addresses these limitations by combining both concrete and symbolic execution of a program, executing it with both concrete and symbolic values simultaneously \cite{Xiao, Sen, Godefroid_DART, Baldoni}.

\begin{figure}
		\begin{center}
			\includegraphics[width = 6cm , height = 6 cm, keepaspectratio]{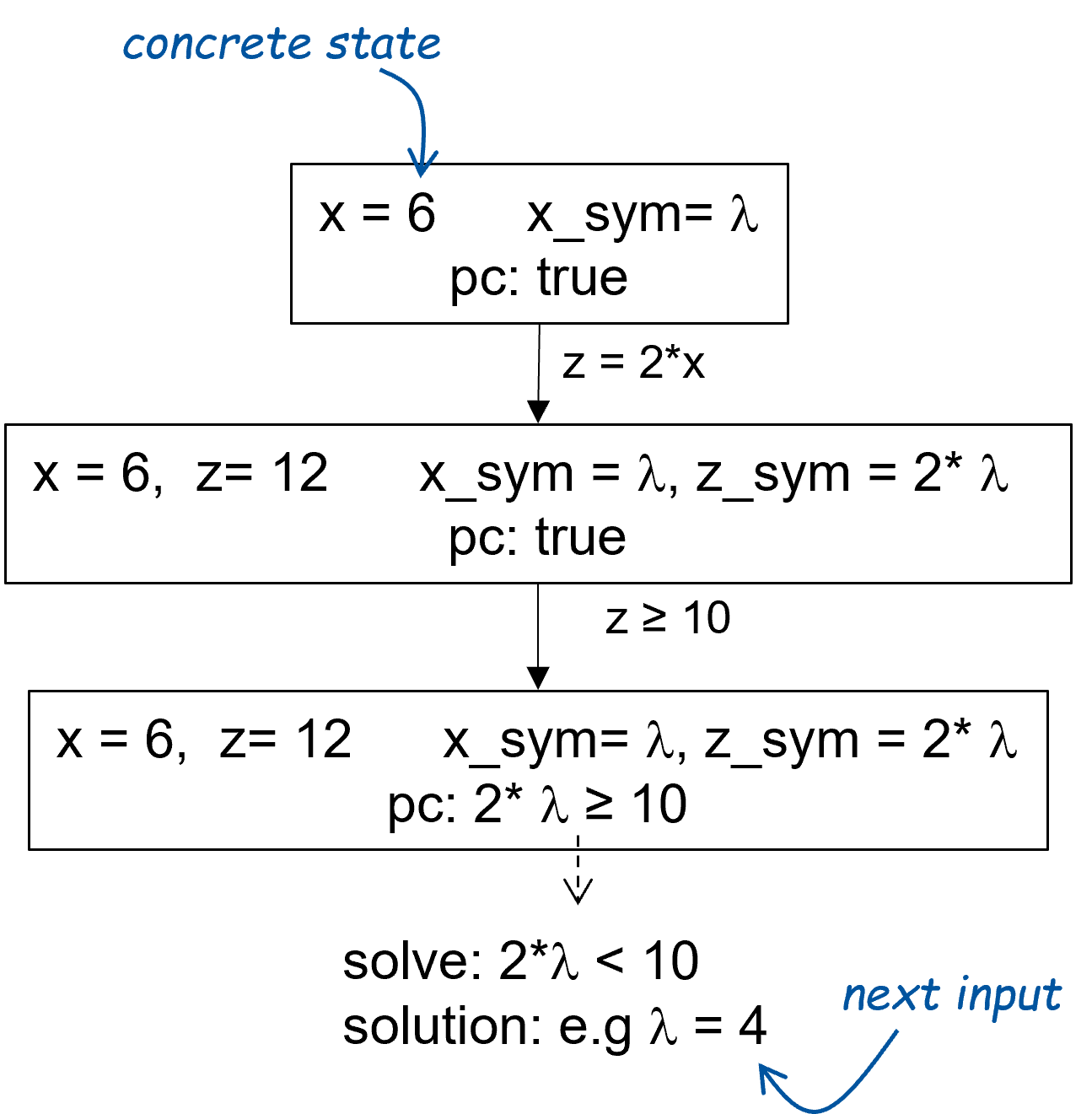}
			\caption{Representation of a specific path executed through Dynamic Symbolic Execution with input $x = 6$}
			\label{dynamic_symbolic_execution_tree}
		\end{center}
	\end{figure}

Referring back to our example program, we can arbitrarily choose a concrete value for the input, such as $x=6$. This choice leads to a specific execution path, as illustrated in \cref{dynamic_symbolic_execution_tree} on the right side. Given that the concrete value for the variable $z$ is calculated to be $12$, the condition \texttt{z < 10} is not satisfied, resulting in the termination of this particular execution path.

When comparing the symbolic path information gathered from both symbolic execution and DSE, we find that there are no significant differences in the information collected. To explore new paths using DSE, the current path constraint is negated, and this modified constraint is then passed to an SMT solver. In our example, the corresponding formula would be $x*2<10$. If this formula is satisfiable, the solver will produce a model containing concrete values that satisfy the condition. In this case, the solver might yield $x=4$.

Further details regarding the design decisions related to symbolic execution and DSE for MontiArc are elaborated in \cref{subsec:design}.
	
    \section{Related Work} \label{sec:relatedWork}

Symbolic execution is a powerful technique used in program analysis that systematically explores program paths by treating inputs as symbolic variables. However, it is computationally intensive, and while advancements in modern computer hardware have facilitated the development of efficient theorem provers and scalable analysis tools, the challenges inherent in symbolic execution have prompted renewed interest in this area \cite{Godefroid_SMART}.

Despite its strengths, symbolic execution faces two primary limitations: path explosion and the complexity of constraint solving. These challenges are also prevalent in Dynamic Symbolic Execution (DSE) and significantly hinder scalability. For instance, constructs such as loops can lead to an exponential increase in the number of feasible execution paths, making it infeasible to systematically analyze all possible paths in large systems or programs. Additionally, the second limitation pertains to constraint solving, which is an NP-hard problem. In larger systems, path constraints can involve intricate combinations of conditions, further complicating the analysis. Consequently, achieving complete path coverage remains elusive in practice \cite{Godefroid_DART, Godefroid_Fuzz, Baldoni, Williams}.

To mitigate these limitations, several approaches have been proposed and integrated into various tools. Notable examples include Microsoft's SAGE, CUTE, DART, and SMART, each designed to enhance the efficiency and effectiveness of symbolic execution in addressing the challenges of path explosion and constraint complexity.

CUTE \cite{Sen}, a concurrent unit testing engine, can efficiently explore paths in C code, achieving high branch coverage and bug detection. 
Due to the limitations of DSE with respect to path explosion, an execution strategy has been used to counteract this. 
CUTE uses a bounded depth-first search to counteract an infinite exhaustive search of the entire computation tree.
Three optimizations have been implemented to counteract the limitations of constraint solving: first, the \textit{fast unsatisfiability check}, which checks whether the last constraint of the path condition is syntactically the negation of any previous constraints, second, \textit{common sub-constraints elimination}, in which the solver identifies and eliminates common arithmetic sub-constraints, and third, \textit{incremental solving}, where the solver identifies dependencies between sub-constraints and exploits them to solve constraints faster and keep solutions similar. 
These optimizations reduce the number of sub-constraints and thus optimize the runtime.
    
jCUTE \cite{jCute} was one of the pioneering symbolic execution tools designed specifically for Java programs. However, as jCUTE is no longer actively maintained, JDart has emerged as a robust replacement \cite{jDart20}. JDart employs DSE to analyze Java programs, primarily focusing on assertion checking. Through this approach, JDart can either identify assertion violations, exhaustively explore all program paths, or reach resource limits during analysis.

One of the key advantages of JDart is its capability to handle complex software systems, as exemplified by its development to analyze intricate NASA software. JDart's architecture is modular, comprising two primary components: the \textit{executer} and the \textit{explorer}. The \textit{executer} is responsible for executing the program and collecting symbolic constraints, leveraging the Java PathFinder framework \cite{havelund2000model, visser2004test}. In contrast, the \textit{explorer} determines the search and execution strategy employed during analysis.

JDart was designed with two main objectives in mind. The first is to create a robust framework capable of managing industrial software challenges, such as handling crashes and addressing the constraints of constraint solving. The second objective is to establish a modular and extensible platform, allowing for the interchangeability of components, the use of various constraint solvers, and the implementation of multiple search strategies or termination conditions \cite{jDart16}.

SAGE \cite{Godefroid_Fuzz}, developed by Microsoft, is an automated white-box fuzz testing tool widely utilized for detecting bugs and vulnerabilities in applications, such as those running on Windows. SAGE implements a technique known as \textit{Generational Search}, which minimizes redundancy while maximizing the generation of new test cases. This systematic approach allows for effective analysis of the state space, enabling SAGE to manage large applications with substantial input sizes. Additionally, SAGE employs heuristics to enhance code coverage, further improving its effectiveness in identifying potential issues.
    
Another tool developed by Godefroid is DART \cite{Godefroid_DART}, which focuses on automatic software testing using DSE. Building on DART's capabilities, SMART \cite{Godefroid_SMART} introduces the concept of compositional analysis, wherein composite functions are decomposed into atomic functions for individual analysis. The results of these analyses are then synthesized through the use of pre- and post-conditions.

While many of the tools discussed earlier primarily focus on program analysis, our research is concerned with a different form of semantic model analysis. Notably, these tools typically employ interpretation rather than generating symbolic code. In our case, we opted to extend the existing code generator for MontiArc rather than develop a new interpreter from the ground up. Similar to JDart, our approach incorporates a modular architecture, separating the execution and collection of symbolic constraints from the implementation of search and execution strategies. Since our implementation is in its initial version, we have yet to incorporate optimizations to mitigate the challenges of path explosion and constraint solving; these enhancements are planned for future work.

A key aspect that differentiates our research from previous work on DSE is our intended goal: to ascertain the semantic differences between executable component-and-connector architecture models. As noted in \cref{sec:introduction}, semantic differencing operators have been developed for various modeling languages \cite{MRR11a, MRR11b, DKMR19, DEKR19, Kau21}, and component-and-connector structures, such as statechart systems (SCS) or automata, are no exception \cite{BKRW17, BKRW19}. However, these approaches typically rely on translations to Büchi automata, which necessitate a finite state space and a defined input-output alphabet. Additionally, to compare entire architectures, syntactic composition of the automata is required, posing challenges when feedback loops are present. Our DSE-based approach, in contrast, circumvents these limitations.
	

\section{Running Example} \label{sec:caseStudy}

    \begin{figure*}[h!]
		\begin{center}
			\includegraphics[width = 0.9\textwidth, keepaspectratio]{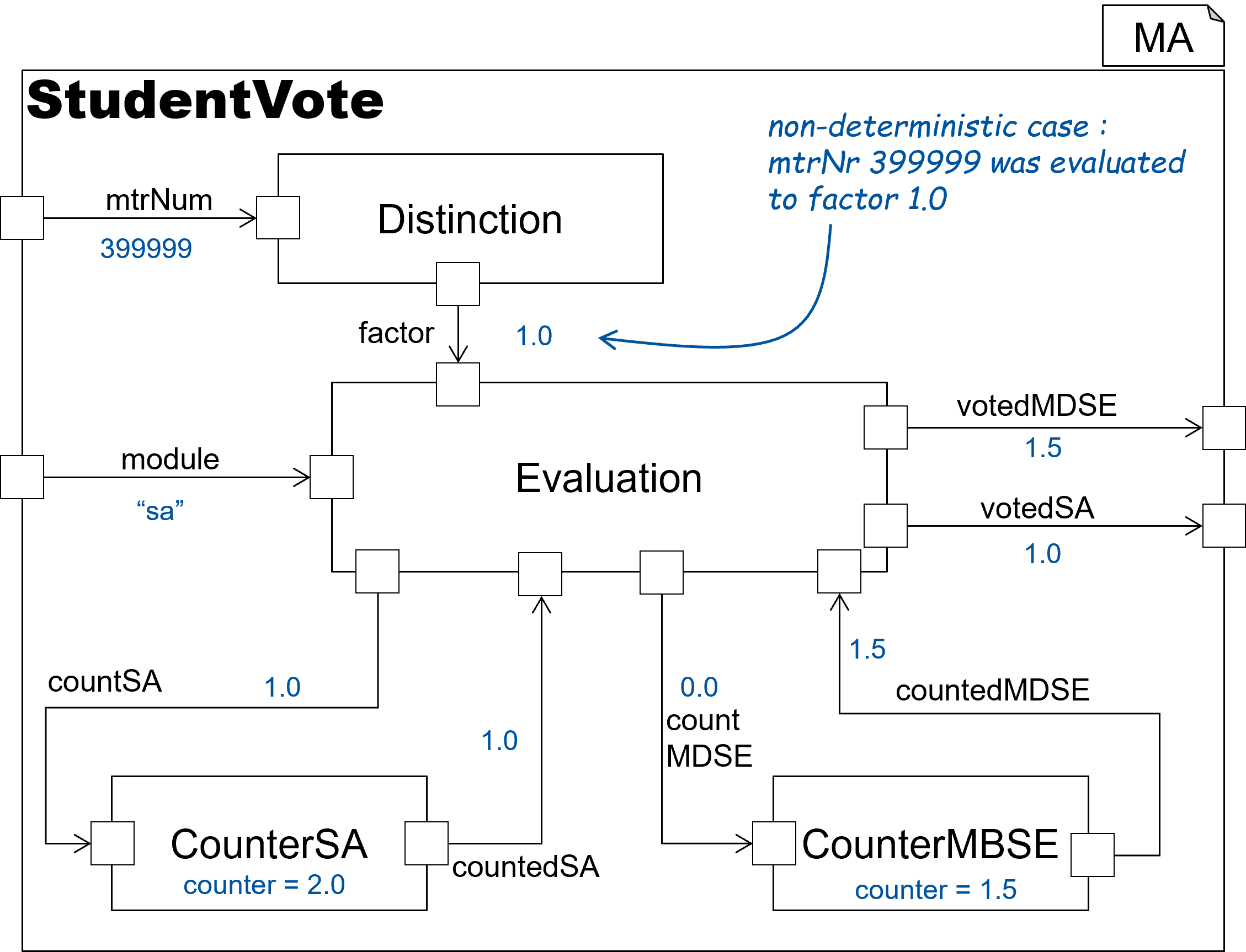}
			\caption{Architecture description of the \texttt{StudentVote}, with the representation of the internal state}
			\label{smallModel_exampleRun_ma}
		\end{center}
	\end{figure*}

MontiArc~\cite{HRR12,Hab16,BKRW17a} is an architectural description language specifically designed for component-and-connector architectures. In this framework, components can either be atomic or consist of sub-architectures, allowing for a hierarchical organization of system elements. Communication between these components is achieved through a message-oriented approach, utilizing channels that connect output and input ports. The input-output behavior of atomic components is described by automata that are a variant of \scs~\cite{statechart,Rum16} based on the FOCUS formalism~\cite{BS12}.

These automata facilitate underspecification due to partiality—where certain inputs may have missing transitions—and non-determinism, characterized by the existence of multiple transitions for the same input at a given state. In the following sections, we introduce a MontiArc model that will serve as the running example throughout this paper.

In \cref{smallModel_exampleRun_ma}, we illustrate the architecture of the MontiArc model named \texttt{StudentVote}. The primary objective of this system is to survey university students and identify the most popular course from two available options: \enquote{Model-Based Systems Engineering} (MBSE) and \enquote{Software Architecture} (SA). The \texttt{StudentVote} model is a composite structure that comprises four sub-components. The first sub-component, \texttt{Distinction}, calculates a weight for each student’s vote based on their matriculation number. Older students, represented by lower matriculation numbers, are assumed to possess a more informed opinion; thus, their votes are assigned a weight of $1.5$. Conversely, younger students, indicated by higher matriculation numbers, receive a weight of 1.01.0. The parameter defining this distinction within the \texttt{Distinction} component can induce non-deterministic behavior if set above $350000$. Specifically, if a student's matriculation number falls between $350000$ and the specified parameter, the weight of their vote can randomly be either of the two factors: $1.0$ or $1.5$.
    
The \texttt{Evaluation} sub-component is the second element in the \texttt{StudentVote} model. It receives the \texttt{factor} corresponding to the student's vote and forwards this value to the appropriate counter module. To cast a vote, the input must be either \texttt{"mbse"} or \texttt{"sa"}; any other input will result in the vote being disregarded. The two counters are interconnected within a feedback loop that includes the \texttt{Evaluation} component. To mitigate issues arising from the sequential processing of models, a delayed connection is implemented, allowing for effective feedback within the composite model. Each delayed port is assigned an initial value, and the resulting input value is stored for use in the subsequent processing step. Consequently, only multiple consecutive inputs can yield differing outputs. Notably, the first output generated by the \texttt{StudentVote} model is consistently 0.00.0, irrespective of the input provided.

In the following a possible execution, with input length 3 will be discussed.
Initialization is done with parameter value $400000$, creating a non-deterministic range between matriculation numbers $350000$ and $400000$.
The first input is \texttt{(355555, "mbse")}, belonging to a student with matriculation number $355555$ and vote for module MBSE.
The delayed ports in the counter components, lead to output \texttt{(0.0, 0.0)}.
The counter for module MBSE is displayed first, followed by the counter for module SA. However, the internal state of the model shows that a factor of $1.5$ was assigned.
Hence, \texttt{CounterMBSE} is incremented accordingly.
The next student might generate the following input \texttt{(500000, "sa")}.
The corresponding output \texttt{(1.5, 0.0)} represents the state after the first input.
The internal state, however, differs, as the \texttt{CounterMBSE} is still $1.5$ and \texttt{CounterSA} was incremented by $1.0$
After the third input message, the output would be the state of the second input.
A third input consisting of \texttt{(399999, "sa")} results in another non-deterministic case.
This time, the factor $1.0$ was assigned, resulting in an increase of the \texttt{CounterSA} by $1.0$.
The output in step three \texttt{(1.5, 1.0)} shows the state after the second input.

The running example demonstrates how branches such as $mtrNum<400000$ and non-deterministic behavior are handled.
Additionally, the integer-values show that the approach works on large state-space. 

    \section{Design and Concept}\label{sec:concept}
This section discusses the main design decisions and provides a high-level overview of the developed functionalities and the implemented controller. Finally, we apply our Dynamic Symbolic Execution (DSE) approach to compute the semantic differences between two MontiArc models.

The implementation of the tool is publicly available as part of the MontiArc project on GitHub\footnote{\url{https://github.com/MontiCore/montiarc}}

\subsection{Major Design Decisions} \label{subsec:design}
This subsection outlines the key design decisions made during the development of the tool.

\paragraph{Dynamic Symbolic Execution vs. Symbolic Execution}
As discussed in \cref{sec:introduction}, DSE is a specialized variant of symbolic execution. We opted for DSE over traditional symbolic execution for two primary reasons.

First, when using symbolic execution, a program's analysis generates a tree structure containing all possible path combinations. This may lead to paths that cannot be satisfied by any concrete values. For instance, consider a function with nested \texttt{if} statements that have mutually exclusive conditions. Take, for example, the conditions \texttt{z < 10} and \texttt{z > 10}. In this case, one of the possible paths would require satisfying both conditions simultaneously, which is inherently impossible.

In contrast, DSE avoids constructing such infeasible paths, as it inherently recognizes that no concrete values can satisfy the path conditions.
    
The second reason for choosing DSE is that not all functions can be effectively analyzed using traditional symbolic execution. For instance, cryptographic functions, such as hash functions, cannot be evaluated through symbolic execution without undermining their security properties. In the case of DSE, these functions can be analyzed by utilizing concrete values for variables. By substituting concrete values, we can evaluate the function and consequently derive new input values for further analysis \cite{Godefroid_SMART}. In summary, DSE was selected because it enables the analysis of functions that are not amenable to traditional symbolic execution and avoids exploring paths with unsatisfiable conditions.

\paragraph{Z3 as an SMT Solver}
Z3 is an award-winning SMT solver developed by Microsoft. It is primarily used for predicate abstraction, advanced static checking, and test case generation. Interaction with Z3 can occur through the SMT-LIB format as well as various APIs \cite{efficientZ3}. Our implementation utilizes the Java API for communication with the solver. Z3 supports multiple theories, including equality, arithmetic operations, and uninterpreted functions. Additionally, it generates a model containing concrete values for all defined constants within the formula. Z3 also includes numerous optimizations and tactics to enhance its performance \cite{bjorner2019programming}.
	
\paragraph{Supported MontiArc Features}
In our approach, we currently limit the behavior descriptions of MontiArc models to automata that incorporate the following features: Ports and internal variables of the automata can be of primitive types, strings, or enumeration types. Additionally, the composition of models, cyclic connector loops, and the definition of model parameters are supported. Notably, our developed extension of the generator also facilitates non-deterministic transition selection, which was previously unsupported.

\paragraph{Generation instead of Interpretation}
Unlike other DSE engines that perform Dynamic Symbolic Execution by executing the program under test and collecting symbolic information without explicit code generation \cite{Godefroid_Fuzz, jDart20, Sen}, MontiArc employs a code generator that produces executable Java code \cite{Hab16}. In this paper, we evaluate a DSE approach based on this existing code generator and extend it to generate symbolic code. We also discuss alternative methods, such as using an interpreter, in \cref{sec:discussion}.
	
	\subsection{Overview of the Tool's Functionalities} \label{subsec:functionalities}
    
\begin{figure}
		\begin{center}
			\includegraphics[width = .8\textwidth, keepaspectratio]{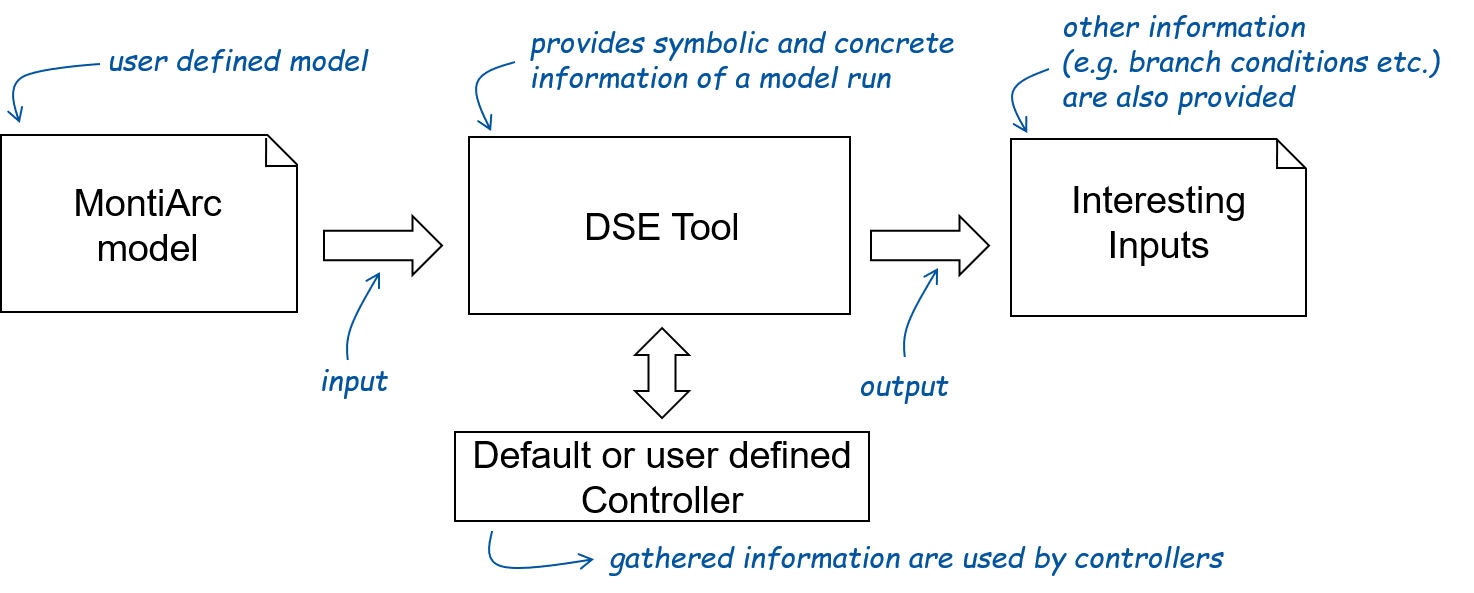}
			\caption{High level overview of the developed tool}
			\label{highlevelOverview}
		\end{center}
	\end{figure}

	The input of the generator is defined as a MontiArc model. 
    This model may only contain the supported MontiArc features, listed in \cref{subsec:design}. 
    MontiArc models are converted into symbolic Java code using the aforementioned extended generator.
    During the execution of the generated symbolic code, a variety of information is collected, including transition condition, taken transitions and state information.
	State information of an automaton can be considered with or without the symbolic and concrete values of internal variables. 
    Collected information is transmitted to a controller, which defines the overall output of the tool and \emph{interesting inputs}. 
    Interesting inputs are input-output pairs and their corresponding branch conditions. 
    Branch conditions represent the specific path taken, as they are the transition conditions satisfied by the corresponding input.
    A high level overview of the tool is displayed in \cref{highlevelOverview}.
    In the following the main functionalities and features are presented.
	
\paragraph{Symbolic Java Code}
Collecting and analyzing information requires symbolic Java code. This code is generated by the extended code generator, with a significant enhancement being the ability to process both symbolic and concrete values of variables simultaneously. To achieve this, ports and internal variables are represented using a new data type called \texttt{AnnotatedValue} (see \cref{annotatedValue}). The \texttt{AnnotatedValue} class features two attributes that represent the variable's symbolic and concrete values.

To extract the information encoded within \texttt{AnnotatedValue}, a second modification was made to the generator. In previous implementations, transitions were converted into if-queries, where the condition represented the transition constraint. To systematically capture transition information, transitions are now generated as follows: for each transition, the function \texttt{TestController.getIf()} is invoked. This function takes the concrete values of all required variables, a boolean expression representing the symbolic condition, and the transition name as inputs. The controller then determines whether the transition can be executed and stores all relevant transition information. An example of such a generated if statement for a transition related to the matriculation number value is shown in \cref{ifQuery}.

Additionally, information about the current state of the model is collected at the end of each execution and passed to the controller for further processing.

\begin{lstlisting}[language = java, float, caption={Java class that stores symbolic expression in addition to value},label=annotatedValue]
public class AnnotatedValue<SMTExpr extends Expr<? extends Sort>, T>{
  private final SMTExpr expr;  // From Z3 SMT Solver
  private final T value;       // Original Message
    //...
}
\end{lstlisting}

\begin{figure*}[h]
		\begin{center}
			\includegraphics[width = 0.9\textwidth, keepaspectratio]{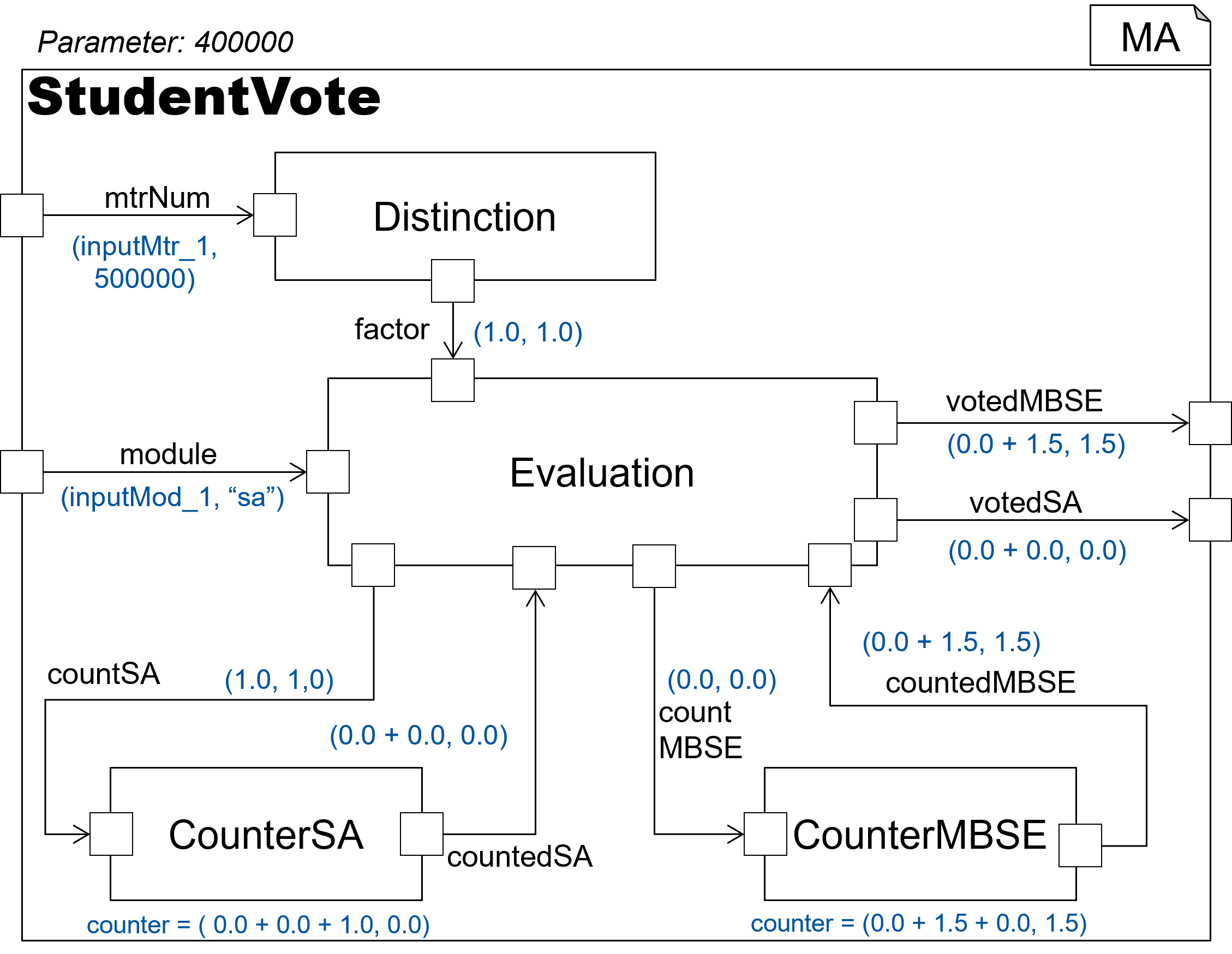}
			\caption{Symbolic and concrete representation of \texttt{StudentVote} internal state after the second input message}
			\label{smallModel_exampleRun_symbolic}
		\end{center}
	\end{figure*}

	\begin{lstlisting}[language = java, float, caption={Exampel of a generated if-query for a transition},label=ifQuery]
BoolExpr expr = ctx.mkGT(matr.getExpr(), 
                            ctx.mkInt(35000);
   
if(TestController.getIf(expr, 
          mtrNr.getValue() > 350000, "branchId")
   ){...}
//...
}
\end{lstlisting}
In \cref{smallModel_exampleRun_symbolic}, the model \texttt{StudentVote} from our motivating example is presented, with its internal state depicted through both symbolic and concrete values. A student with a matriculation number of $500000$ votes for the module SA. Each internal variable and port is assigned a symbolic and concrete representation; for example, the input is represented as  \texttt{(inputMtr\_1, 500000)} and \texttt{(inputMtr\_1, "sa")}.

For ports, only the current values are visible, as previous values are not retained. In contrast, internal variables, such as counters, may be influenced by prior inputs. The symbolic values of internal variables indicate how the current values were derived.

The expression \texttt{(0.0 + 0.0 + 1.0, 1.0)} represents the current symbolic state of the counter for SA. The initial and first input was 0.00.0, and the current computation has incremented the counter by one.
	
\paragraph{Information Gathered through Analysis}
During the execution of a model, various types of information are collected. This includes all possible transitions starting from the current state of the automaton, as well as all transitions that have been taken. For each transition, the collected information encompasses branch conditions and the unique name of the transition. The branch conditions specify the particular path taken through the model and must be satisfied by the input.

After processing each input message, information about the state of the automaton is gathered. This includes the current state's name, along with the symbolic and concrete values of internal variables. This data is collected for each sub-component and is subsequently combined into the state of the composite model.

All information regarding states and transitions is transmitted to the controller, where it can be utilized to implement termination conditions.
	
\paragraph{Functionality of a Controller}
A controller is responsible for implementing the execution strategy and the logic required for defining new inputs. To facilitate easy integration, each controller must adhere to a specific interface, allowing users the flexibility to implement the details according to their needs. The following paragraphs will elaborate on the general concept of a controller and its execution strategies, such as path coverage.
    
\begin{figure}
		\begin{center}
			\includegraphics[width = 0.7\textwidth, keepaspectratio]{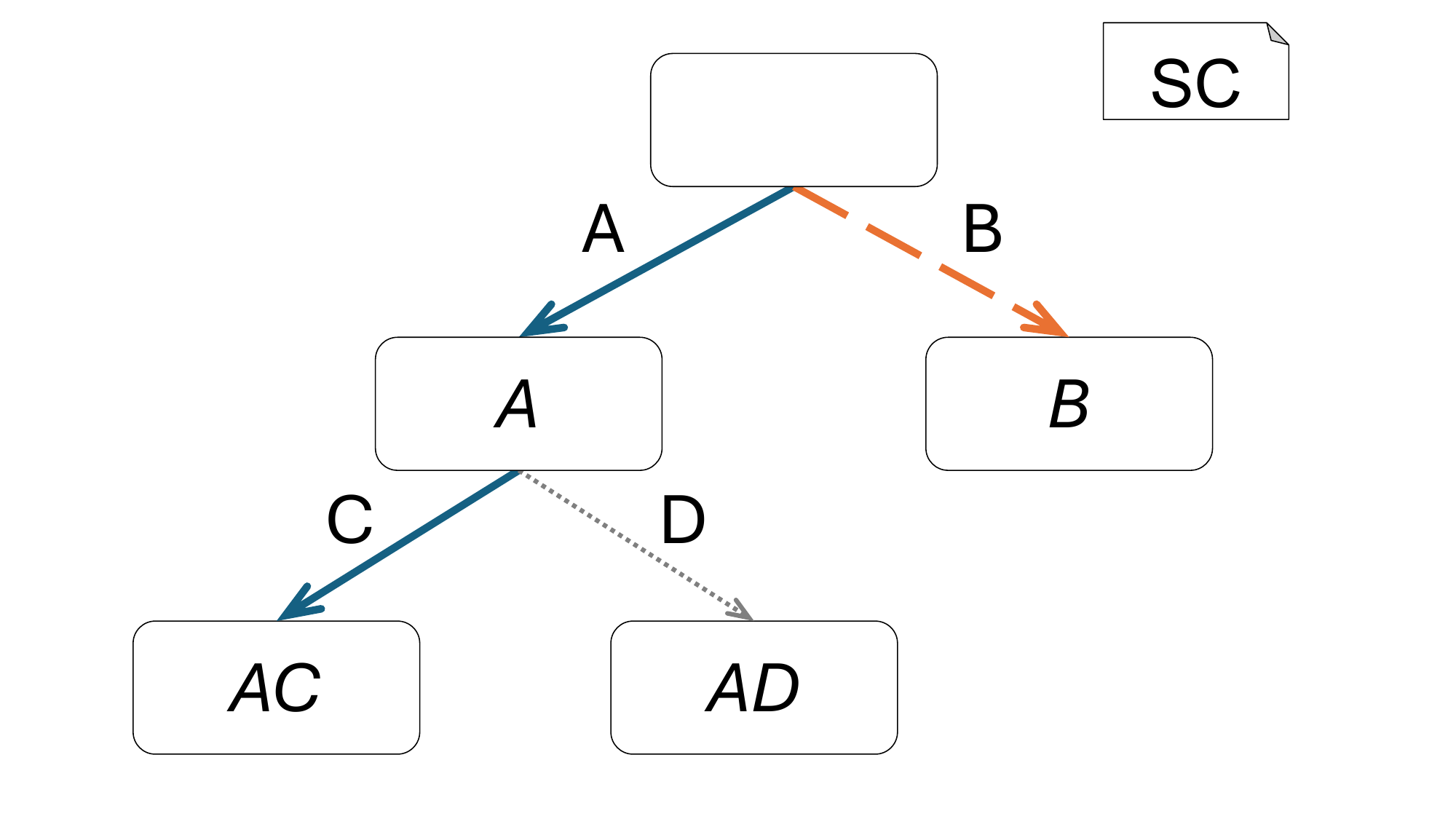}
			\caption{Example of an automaton to illustrate a controller where conditions A and D are satisfied by the input.}
			\label{explanation_controller_b}
		\end{center}
	\end{figure}

In \cref{explanation_controller_b}, an example automaton is presented. During model execution, each branch condition is collected and passed to the controller. Consider an input length of two; the solid blue arrows represents the first path taken. The controller is aware of the following information: \texttt{[A, -B, C, -D]}.

To achieve path coverage, an iterative negation of path constraints is employed. The first constraint is negated and provided to the solver, resulting in a new model that ensures a different path is taken. In the second recursion, the path represented by dashed orange arrows is taken, and \texttt{[-A, B]} is collected. Next, \texttt{B} is negated. Since the combination \texttt{[-A, -B]} is unsatisfiable, the controller returns to the first recursion. There, the next constraint to be negated is \texttt{C}, as \texttt{A} and \texttt{B} were previously negated.
    
It is also possible to implement conditions within a controller to further control the paths taken; for example, ensuring that each transition is visited only once. In this case, the first and second recursions would remain the same as before. However, upon returning to the first recursion and checking \texttt{[A, -B, -C]}, the behavior would differ. Previously, a new path would be found and taken, but this path includes the transition with condition \texttt{A}, which has already been visited. As a result, the path execution is terminated, and the next possible path is explored.

In addition to branch conditions, the controller collects information regarding the state of the automaton. This information can be used to evaluate the controller or to validate the model, such as for path redundancy and non-determinism. These aspects are presented below.

We have developed multiple controllers, categorizing them into the following three groups:

\paragraph{Category 1: Path Coverage}
Controllers in the \texttt{Path Coverage} category aim to discover all possible paths within a model. Variations of these controllers implement oracles to handle non-deterministic models or utilize search algorithms.

\paragraph{Category 2: Termination Condition}
This category includes all controllers that utilize specified termination conditions. These conditions can be based on visited transitions or states, with or without involving internal variables. An example of such a termination condition can be found in \cref{subsec:functionalities}.

\paragraph{Category 3: Random Generation}
Controllers that do not utilize the symbolic information gathered by the tool fall under the category of \textit{Random Generation}. This includes controllers that execute only once based on the given initial input or those that randomly generate input values that conform to the specified input types.
	
\paragraph{Path Redundancy}

Path redundancy arises when two distinct paths in a model exhibit identical simplified branching conditions. In the context of white box testing—where the internal workings and input-output behavior of individual components are analyzed—both paths are significant. Conversely, when the model is approached as a black box, both paths represent the same scenario; thus, only one occurrence is necessary to reflect the behavior of the system.

To identify path redundancy within a model, we examine the simplified representations of the branching conditions and outputs. If these simplified expressions are equivalent, we can conclude that a pair of redundant paths has been identified. For instance, consider the two paths: the first path has a branch condition of \texttt{x + 1 < 5} and an output of \texttt{x + 1 - 2}. The second path consists of the branch condition \texttt{x + 2 < 6} and an output of \texttt{x + 2 - 3}. Upon simplifying these expressions for both paths, we find that they share the same branch condition of \texttt{x < 4} and an output of \texttt{x - 1}. This finding illustrates how path redundancy can simplify the analysis of models, ensuring efficiency in testing and validation processes.
	
\paragraph{Non-Determinism of the Model}
The extended generator supports models that include non-deterministic automata, which were previously unsupported by the code generator. In this context, non-determinism means that for a given input sequence, multiple paths and corresponding output sequences are permissible within the model.

Using the collected information, it is possible to determine whether a model is deterministic. Each path taken, represented by a set of branch conditions, is decomposed into symbolic expressions while preserving the order. For example, consider the following two paths:

The first path is represented by the condition \texttt{x > 5 $\land$ y + 2 < 6}, where \texttt{x} and \texttt{y} are symbolic constants representing the input. The second path is represented by \texttt{x > 5 $\land$ y + 2 < 8}.

Decomposing both paths yields the following representations:

\begin{center}
\texttt{[x > 5, y + 2 < 6]} and \texttt{[x > 5, y + 2 < 8]}.
\end{center}
    
To assess the non-determinism of two paths, we compare the symbolic expressions at each corresponding position in their respective expression lists. Initially, we check each pair of expressions for equality. If they are found to be unequal, we then evaluate their satisfiability when combined through conjunction. If all pairs either contain equal expressions or yield satisfiable results when combined, we conclude that the two paths represent non-deterministic alternatives for some input values. Conversely, if any pair is unsatisfiable, it indicates that the paths are completely disjoint concerning the input.

In our specific example, the first pair \texttt{(x > 5, x > 5)} consists of equal expressions, while the second pair \texttt{(y + 2 < 6, y + 2 < 8)} comprises expressions that are satisfiable in conjunction. Consequently, these paths can be considered non-deterministic alternatives for certain values of \texttt{x} and \texttt{y}.

By employing this pairwise comparison approach across all paths, we can ascertain the number of non-deterministic paths. However, this method is computationally demanding and does not scale well with larger models, making it impractical for extensive applications. Furthermore, using an oracle to identify non-deterministic occurrences is not feasible, as such information is not mandatory. To address this limitation in larger models, we focus solely on detecting the presence of non-deterministic paths. Although this approach is less computationally intensive, it still necessitates multiple solver calls and comparisons, which can result in prolonged runtimes. Strategies for optimizing solver call usage to mitigate this issue are discussed in \cref{sec:discussion}.

\subsection{Computation of Semantic Differences Between MontiArc Models}
\label{semanticDifference}
	
With the implementation of DSE in MontiArc, we can now compute semantic differences between two MontiArc models. To illustrate this process, we first introduce a modified version of the existing model \texttt{StudentVote}, referred to as \texttt{StudentVoteAlt}. Subsequently, we will explain how to compute semantic differences in the form of diff-witnesses.
	
\begin{figure*}[h]
		\begin{center}
			\includegraphics[width = 0.9\textwidth, keepaspectratio]{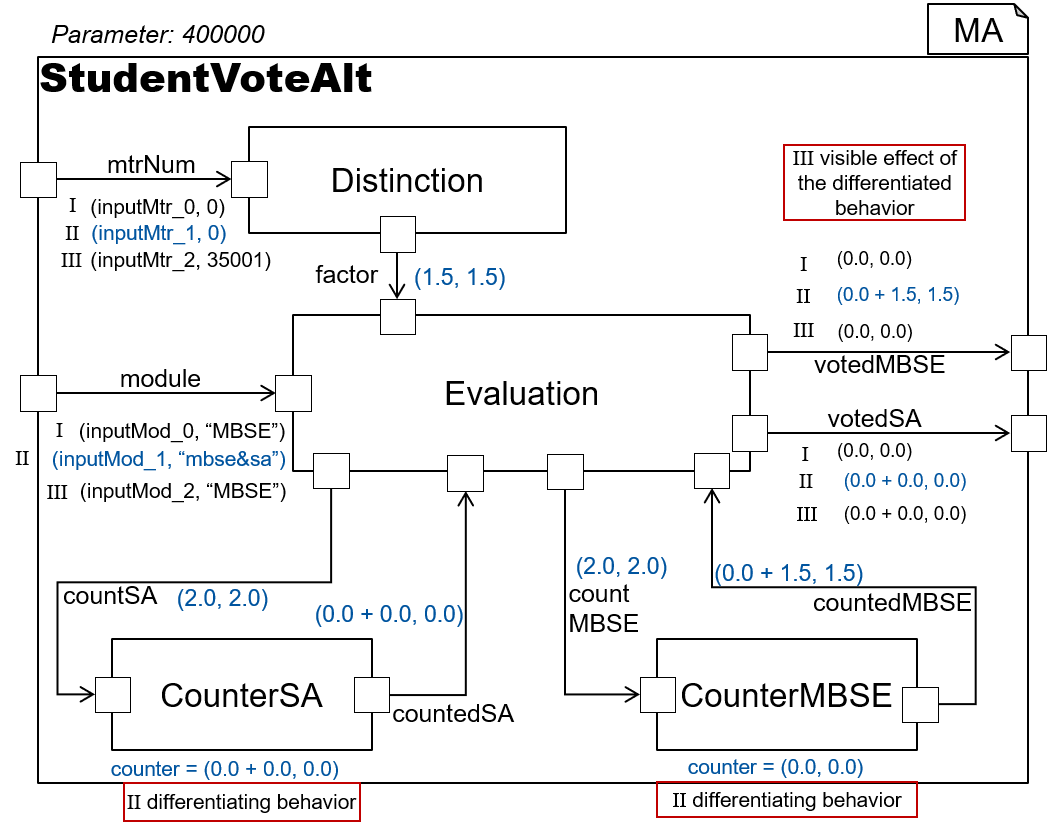}
			\caption{Example of semantic difference between \texttt{StudentVoteAlt} and \texttt{StudentVote}, input length 3 }
			\label{semDiff_example}
		\end{center}
	\end{figure*}

The model \texttt{StudentVoteAlt}, as shown in \cref{semDiff_example}, retains the basic structure of \texttt{StudentVote} but introduces modifications to the \texttt{Evaluation} and \texttt{Counter} components. Notably, \texttt{Evaluation} now allows simultaneous voting for both modules through the message \texttt{mbse\&sa}, resulting in each counter being incremented by $2.0$. 
This addition creates a new execution path that is absent in the original \texttt{StudentVote} model.

Furthermore, each counter component is designed to reset to $0$ once its internal value reaches $1.0$. After this reset, the counter continues to count arbitrarily. Consequently, semantic differences are only observable when the input length exceeds 2. The relationship between input and output—specifically, the diff-witnesses between these two models—is illustrated in \cref{semDiff_example}.

In this comparison, individual inputs are provided to each model for execution. For the first input, both models respond identically. However, the difference arises with the second input. When voting for \texttt{mbse\&sa}, the \texttt{StudentVoteAlt} model determines the increment factor for its counter to be $2$. At this point, the value of \texttt{CounterSA} remains unchanged, as it is less than $1.0$. In contrast, \texttt{CounterMBSE} has a value of $1.0$, triggering a reset to $0$. This discrepancy will only manifest after the subsequent input, thus necessitating an input length of three to observe the difference. This example highlights a path that cannot be replicated in the original \texttt{StudentVote} model.
 
When assessing the semantic difference between two models, it is crucial to consider certain specific scenarios. One such scenario arises when a component in the second model contains an automaton that is deemed partial concerning the possible inputs of the first model. In this case, the partial automaton will respond to any unknown input by \textit{ignoring} it—meaning it neither produces a new output nor transitions to a different state \cite{Rum96}. Consequently, the automaton retains its previous state and output. However, if the output ports are delayed, \texttt{Null} values may be generated. It is essential to address these cases within the automaton's definition of the component behavior.

The semantic difference is computed by comparing the outputs of both models while treating each model as a black box, thus eliminating the need to consider path conditions. The process is outlined in \cref{algorithmSemDiff}. The first model undergoes analysis through DSE, from which input-output pairs are extracted. These pairs are then utilized as inputs for the second model. The simplified symbolic portion of the computed output from the second model is compared to the corresponding output of the first model. If discrepancies are found between the simplified outputs, the input-output pair is classified as a diff-witness.
    
Until this point, we have implicitly assumed that the models under consideration are deterministic. However, our approach is also applicable to non-deterministic models. The non-determinism present in the first component will be managed through DSE. In contrast, addressing the non-determinism in the second component requires an additional step following the initial output comparison. In this scenario, all potential paths must be explored to ensure the soundness of the differencing operator.

To achieve this, we utilize an oracle that represents or determines the non-deterministic \enquote{choices} within the model. All possible oracle values are computed and applied to the specific input. Each resulting output is then compared against the output of the first model. If a matching output is identified, the search is halted, and the comparison of output pairs continues. Conversely, if no matching output is found, the corresponding input-output pair from the first model qualifies as a valid diff-witness. Given that all potential paths through the model have been examined, the calculation of the semantic difference is deemed sound.

\begin{algorithm}
\caption{Algorithm to calculate semantic differences between MontiArc models}
\label{algorithmSemDiff}
\begin{algorithmic}
\State \textbf{Input:} m1, m2
\State \textbf{Output:} List of semantic differences
\Statex
\State $diff\_cases \gets [\:]$
\State $dse\_result \gets execute\_DSE(m1)$
\State $found\_witness \gets$ \textbf{False}
\Statex
\For{\textbf{each} $in\_m1, out\_m1$ \textbf{in} $dse\_result$}
    \State $out\_m2 \gets runModel2(in\_m1)$
    \State $s\_out\_m1 \gets simplify(out\_m1)$
    \State $s\_out\_m2 \gets simplify(out\_m2)$
    \Statex
    \If{$s\_out\_m1 \neq s\_out\_m2$}
        \State $found\_witness \gets$ \textbf{True}
        \State $oracles \gets calc\_oracles(m2, in\_m1)$
        \Statex
        \For{\textbf{each} oracle \textbf{in} oracles}
            \State $out\_oracle \gets runModel2(in\_m1, oracle)$
            \State $s\_out\_oracle \gets simplify(out\_oracle)$
            
            \If{$s\_out\_m1 == s\_out\_oracle$}
                \State $found\_witness \gets$ \textbf{False}
                \State \textbf{break} 
            \EndIf
        \EndFor
        \Statex
    \EndIf

    \If{$found\_witness$}
        \State $diff\_cases \gets diff\_cases \cup \{in\_m1, out\_m1\}$
    \EndIf
\EndFor
\Statex
\Return $diff\_cases$

\end{algorithmic}
\end{algorithm}
	\section{Evaluation} \label{sec:evaluation}
In this section, we evaluate the tool and the implemented controllers using the \texttt{StudentVote} example introduced in \cref{sec:caseStudy} as our case study. We will first define the evaluation criteria, followed by a presentation of the results, including the calculation of semantic differences between two MontiArc models.

\subsection{Definition of Evaluation Criteria and Controllers}
We have established the following criteria for evaluating the implemented controllers: \textit{Runtime}, \textit{Minimality}, and \textit{Completeness}.

\paragraph{Runtime}
The runtime of the developed tool plays a crucial role in determining its usability. To facilitate comparison, the runtime of each controller is classified into one of the following categories:
\begin{itemize}
    \item \textit{Constant Runtime}: The runtime remains unchanged regardless of the input length.
    \item \textit{Linear Runtime}: The runtime increases proportionally with the input length.
    \item \textit{Exponential Runtime}: The runtime grows exponentially as the input length increases.
\end{itemize}

Among these categories, controllers exhibiting constant runtime are deemed most desirable, followed by those with linear runtime, while exponential runtime is considered the least favorable option.
	
\paragraph{Minimalism}
The set of interesting inputs collected during the analysis should be minimized to ensure that the outputs remain simplified. Due to Z3's non-deterministic behavior in selecting concrete values for a given formula, we focus exclusively on symbolic values. However, simplification of each symbolic output results in the loss of information regarding the paths traversed, effectively treating the model as a black box.

For instance, consider two interesting inputs characterized by the following properties: the first input has a symbolic output of \texttt{x + 1 - 4} and a branch condition of \texttt{x + 1 < 5}, while the second input has a symbolic output of \texttt{x - 5 + 2} and a branch condition of \texttt{x - 5 < 10}. Here, \texttt{x} represents the symbolic representation of the input. Both inputs were derived from different paths, but when both symbolic outputs are simplified, they yield \texttt{x - 3}. Consequently, these two inputs would be considered duplicates concerning the \textit{minimality} criterion.

As a metric, \textit{minimalism} assesses the number of duplicates in relation to the size of the set of interesting inputs. Controllers are evaluated based on the percentage of duplicates, with a higher proportion of duplicates being regarded as less favorable.
	
\paragraph{Completeness}
When employing automata as the behavioral description of components, the completeness of a controller can be assessed from multiple perspectives. One perspective focuses on the number of visited transitions, represented as the ratio of visited transitions to the total number of existing transitions. A similar approach is used for assessing the completeness of states, which is determined by the ratio of visited states to the total number of existing states.

There are two perspectives regarding states: the first considers states without any information about the automaton's internal variables, while the second incorporates the current values of these internal variables as part of the state representation.

Given that the state space is often infinite concerning internal variables, it is not always feasible to calculate the percentage of visited states. Therefore, in such cases, the absolute number of visited states and transitions is used as an alternative metric for completeness.

\subsection{Results of the Evaluation}

In this subsection, we present the evaluation of the controllers using the model \texttt{StudentVote} from our motivating example as a case study. The implications of these results are discussed in \cref{sec:discussion}. It is important to note that all calculations and runtime measurements are specific to the hardware utilized during the evaluation. The evaluation was conducted on a ThinkPad T14s equipped with 32GB of RAM, an AMD Ryzen 7 Pro processor running at 2.7 GHz, and featuring 8 cores.

\begin{figure*}
		\begin{center}
			\includegraphics[width=\textwidth,keepaspectratio]{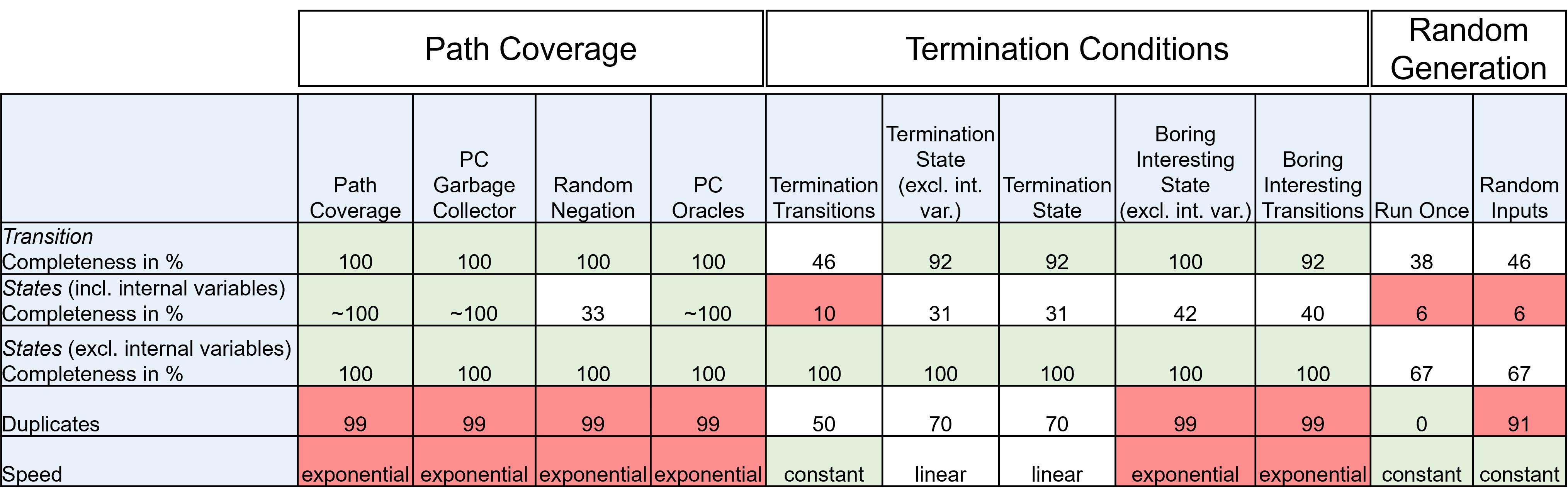}
			\caption{Results regarding the \texttt{StudentVote} based on input length 4}
			\label{result_evaluation_smallModel}
		\end{center}
	\end{figure*}

The detailed results of the evaluation are presented in \cref{result_evaluation_smallModel}. In this visualization, poor performance is indicated in red, while good performance is marked in green. The results for the categories \texttt{Runtime} and \texttt{Completeness} are classified as follows: a percentage below 25\% is deemed poor, while a percentage above 75\% is considered good, with values in between categorized as neutral. Conversely, in the category \texttt{Minimalism}, results above 75\% are regarded as deficient, whereas those below 25\% are classified as proficient.
	
\paragraph{Runtime}
The controllers categorized under \texttt{Path Coverage} experience exponential runtimes. For instance, an input length of 6 is estimated to take approximately 1.9 days to process. This exponential behavior can be attributed to the number of solver calls, which are computationally intensive and contribute significantly to the overall runtime, even for relatively small models and short input lengths.

In contrast, for the category \texttt{Termination Condition}, no uniform runtime can be established. When the termination condition is defined based on transitions, the runtime remains constant. However, when it relies on states—whether internal states are included or not—the runtime becomes linear. An exception arises with the \texttt{Boring Interesting Controller}, which utilizes specific states and transitions for its termination condition, leading to an exponential runtime.

The variability in runtime within this category can be explained by the nature of the termination conditions. For instance, if a condition is based on the frequency of visits to a particular transition in the automaton, the number of recursive calls is limited by that frequency, effectively creating a bottleneck. Similarly, state-based termination conditions can also act as bottlenecks. Moreover, controllers of the \texttt{Boring Interesting} type may suffer from poorly categorized transitions or states. Ineffectively chosen termination conditions can lead to situations where a transition is erroneously classified as interesting, allowing for numerous visits—even if that transition is never actually executed—resulting in the termination condition failing to apply.
    
Controllers in the category \texttt{Random Generation} exhibit a constant runtime. This consistency can be attributed to the non-utilization of collected symbolic information and the implementation of a cap on the maximum number of identifiable paths, which prevents infinite runtimes. The controller \texttt{RunOnce} executes the component precisely once for the specified input length, while \texttt{RandomInput} explores new paths exactly ten times using randomized inputs.

When assessing controllers based on runtime performance, those in the \texttt{Random Generation} category emerge as the most efficient, followed by those in the \texttt{Termination Condition} category. In contrast, the \texttt{Boring Interesting} controllers and those in the \texttt{Path Coverage} category are the least efficient, as they experience exponential runtimes.
	
\paragraph{Minimalism}
The architecture of the \texttt{StudentVote} system allows for multiple inputs to yield the same output. Initially, irrespective of the input provided, the output consistently appears as zero for both counters. This behavior is attributed to the previously discussed delayed ports. For input lengths of 3, the percentage of duplicate outputs ranges from 86\% to 92\%, while for input lengths of 4, this range increases to 91\% to 99\%. Notably, exceptions arise with the controllers \texttt{Termination Transition}, \texttt{Termination Automaton State}, \texttt{Termination State}, and \texttt{RunOnce}. The \texttt{RunOnce} controller uniquely identifies a single path within the model, resulting in the absence of duplicates, regardless of input length. In contrast, the other exceptions can be attributed to a reduced number of interesting inputs discovered due to certain restrictions. Importantly, none of the evaluated controllers employ a strategy to prevent the generation of duplicate interesting inputs in relation to the simplified outputs.
	
\paragraph{Completeness of Visited Transitions}
To achieve completeness, it is essential that each transition is visited at least once. Controllers belonging to the categories \texttt{Path Coverage} and \texttt{Termination Condition} attain a transition coverage of 54\% (7 visited transitions) for an input length of 1. However, a minimum input length of 3 is required to achieve full transition coverage of 100\% (13 visited transitions). The \texttt{Termination Transition} controller is an exception; it is constrained by a limit on the number of visits allowed for each transition, resulting in a bottleneck for the initial two transitions in the \texttt{StudentVote} scenario.

In the case of controllers in the \texttt{Random Generation} category, transition coverage ranges from 31\% to 46\% (RunOnce: 4 or 5 visited transitions; RandomInput: 5 or 6 visited transitions), regardless of input length. This limited coverage is attributed to the restricted number of iterations performed by the controllers, which consequently reduces the number of paths explored.

Overall, all controllers are constrained by the specified input length when it comes to transition coverage. An inadequate input length hinders the complete detection of transitions. Conversely, when a sufficient input length is provided, controllers in the \texttt{Path Coverage} category can achieve full transition coverage. While controllers in the \texttt{Termination Condition} category may have the potential to explore all paths, they are ultimately restricted by their respective termination conditions.
	
\paragraph{Completeness of Visited States Excluding Internal Variables}
Controllers classified under \texttt{Path Coverage} and \texttt{Termination Condition} successfully visit 100\% (6 states) of the reachable states starting from an input length of 1. In contrast, controllers categorized as \texttt{Random Generation} only manage to visit 67\% (4 states) of the states, irrespective of the input length.

\paragraph{Completeness of Visited States}
An automaton's state can encompass the potential concrete values of all its internal variables. Consequently, \texttt{StudentVote} does not possess a finite state space due to its counter components. To quantify the percentage of visited states within the model, we relate it to the maximum number of states that can be reached based on the input length.

Controllers in the \texttt{Path Coverage} category consistently achieve 100\% relative coverage, regardless of input length, except for the \texttt{RandomNegation} controller, which attains only 33\% for an input length of 4. In the \texttt{Termination Condition} category, the percentage of visited states declines as input length increases, resulting in a completion rate of 10\% to 40\% for an input length of 4, largely due to the constraints imposed by their respective termination conditions. Controllers in the \texttt{Random Generation} category achieve merely 6\% coverage for an input length of 4.

The ability to achieve completeness in visited states is contingent on both the model and the specified input length. Shorter input lengths are inadequate for attaining completeness in visited states for larger models.
	
Upon comparing the results, it becomes evident that no single controller emerges as the superior choice. The selection of a controller should be guided by the specific model in use and the objectives of the user. Notably, for controllers categorized under \texttt{Path Coverage}, an exponential increase in runtime is to be anticipated, regardless of the underlying model. If runtime efficiency is a primary concern and only a specific subset of paths is needed, opting for a controller of the \texttt{Termination Condition} type may prove to be the most effective strategy.

Most controllers, particularly those focused on path coverage, experience exponential growth in runtime. This phenomenon is exacerbated by the path explosion, which leads to a corresponding exponential increase in the number of solver calls and solver operations. The average CPU load on a single core due to these SMT-Solver operations is approximately 73 percent, indicating a pressing need for optimization in solver operations. One effective optimization strategy involves implementing a timeout for the solver. If the solver fails to identify a solution within the allotted time, we can infer that the branch condition for the potential path is likely unsatisfiable, allowing it to be skipped.

A critical challenge in this optimization process lies in selecting an appropriate timeout duration. The goal is to maximize time improvement while minimizing result degradation. Here, result degradation is defined as the ratio of interesting inputs identified with the timeout versus those found without it.

To assess the time improvement and result degradation associated with the \texttt{Path Coverage Controller} in conjunction with garbage collection, various timeout durations were tested. The findings are summarized in \cref{timeout_PCCGC}. An optimal balance between time improvement and result degradation appears to be achieved with a timeout set at 10 milliseconds. Shorter timeouts can lead to a degradation of results by as much as 99\%, while longer timeouts yield only a modest time improvement of up to 14\%.

    \begin{table}[htb]
        \centering
        \begin{tabular}{r|r|r}
            Timeout in ms & Time improvement & Result deterioration \\
            \hline
            1 & 0.9970 & 0.9994 \\
            5 & 0.9965 & 0.9988 \\
            7 & 0.98 & 0.97 \\
            8 & 0.88 & 0.83 \\
            9 & 0.70 & 0.59 \\
            10 & 0.15 & 0.03 \\
            30 & 0.14 & 0.00 \\
            300 & 0.18 & 0.00 \\
            1000 & 0.17 & 0.00 \\
        \end{tabular}
		\caption{Different timeouts for input length three, controller: PC Garbage Collector}
		\label{timeout_PCCGC}
    \end{table}
	
In conclusion, implementing a timeout for the solver can lead to significant runtime improvements. However, it is crucial to consider the accompanying loss of results. To achieve the optimal balance between time improvement and result degradation, the timeout must be individually calibrated for each combination of model, controller, and input length.

\paragraph{Computation of Semantic Differences Between MontiArc Models}\label{par:semdiff}
The evaluation of various controllers reveals that no single controller stands out as universally superior. Instead, the selection of a controller should be tailored to the specific needs of the user. Any controller can be employed to calculate the semantic difference; however, if the goal is to compute all diff-witnesses, it is recommended to select a controller from the \texttt{Path Coverage} category. Conversely, if identifying just a single diff-witness is sufficient, a category 2 controller may be more appropriate.

In the subsequent analysis, we compute the semantic difference using the \texttt{Path Coverage Controller} with a garbage collection trigger. The results of the semantic difference analysis between \texttt{StudentVote} and \texttt{StudentVoteAlt} for varying input lengths are presented in \cref{semDiff_result}.
    
    \begin{table}[htb]
        \centering
        \begin{tabular}{r|r|r|r}
            Input length & Time in min & \#Diff-witness & \#Solver calls \\
            \hline
            1 & 0.08 & 0 & 140 \\
            2 & 1.34 & 0 & 2380 \\
            3 & 45.46 & 512 & 37196 \\
            4 & 22227.27 & n/a & n/a \\
        \end{tabular}
    \caption{Results of the computation of semantic differences between \texttt{StudentVote} and \texttt{StudentVoteAlt}}
    \label{semDiff_result}
    \end{table}
    
As anticipated, a diff-witness was only identified starting from an input length of three. The challenges associated with calculating semantic differences mirror those of the selected controller. To analyze larger modules or to accommodate greater input lengths, optimization of the tool is necessary. Various optimization strategies are explored in \cref{sec:discussion}.
    \section{Discussion} \label{sec:discussion}
As discussed in \cref{sec:evaluation}, the runtime of the tool is significantly constrained by the CPU load associated with SMT solver operations. Specifically, the number of server calls increases exponentially with both the length of the input and the size of the model.

To address this issue, we have implemented a timeout strategy for the SMT solver as an optimization approach. The underlying assumption is that if no model is found within a specified timeframe, it is likely that no model exists. This strategy provides a dual benefit: it improves runtime while potentially compromising the quality of the results. Striking an optimal balance between these two factors necessitates selecting an appropriate timeout based on the specific controller in use. An alternative solution could involve parallelizing the analysis; however, the documentation on parallelism is limited, and its implementation for Z3 is not yet complete.

While the strategies presented offer some avenues for improvement, none provide a definitive solution to the runtime limitations. To substantially mitigate this issue, it is essential to reduce the number of solver calls. Achieving this requires the formulation of an effective controller strategy tailored to the specific use case. The challenge lies in minimizing solver calls while maximizing the discovery of paths.

In \cref{sec:evaluation}, we evaluated three categories of controllers, each encompassing various implementations. While each category presents its own set of advantages and disadvantages, none emerged as the unequivocal best option. It is important to note that achieving comprehensive path coverage may necessitate accepting exponential runtime. Conversely, by settling for less exhaustive results, it is possible to attain a linear or constant runtime in relation to the input length.
    
To achieve an optimal balance between completeness and runtime, we propose a combination of controllers. Random input generation yields the least complete results but maintains a constant runtime. In contrast, a path coverage controller offers the most comprehensive results, albeit at the cost of exponential runtime. By integrating both controllers, we can likely attain an acceptable compromise regarding both completeness and efficiency.

\begin{figure}
		\begin{center}
			\includegraphics[width=.8\textwidth,keepaspectratio]{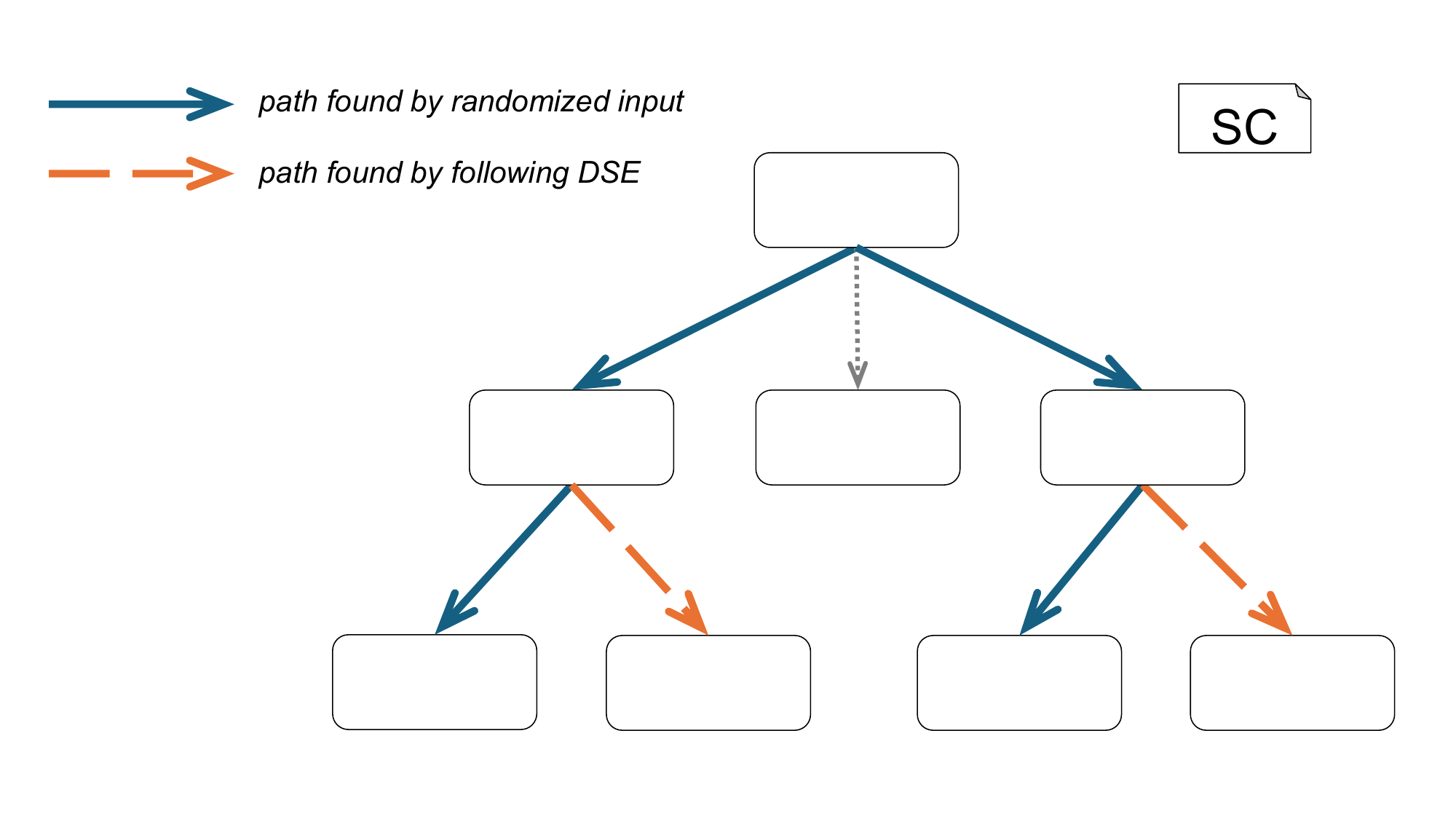}
			\caption{Possible paths found by a controller combining randomized inputs and DSE}
			\label{queens_problem}
		\end{center}
	\end{figure}

One potential combination could be inspired by the queen's problem. Initially, a random input is provided to the model. Utilizing the symbolic information gathered from this input, the system then searches for a defined number of new paths. By generating a new random input and repeating this process, we can facilitate a more thorough exploration of the path tree within a user-defined search range. An example of the paths identified within such a tree is illustrated in \cref{queens_problem}. Future evaluations will be necessary to determine whether this combined controller outperforms the existing implementations.
In order to enable the integration of different controllers, modifications to the current controller architecture are required. This approach holds promise for significantly reducing the runtime of DSE, though the completeness of the results produced by the new controllers will also need to be assessed.

Another avenue for enhancing the tool is through decomposition analysis of models, as introduced by ~\citet{Godefroid_SMART}. This technique involves breaking down the program under analysis into individual functions. The results from analyzing these functions are then synthesized, taking into account their pre- and post-conditions, to evaluate composite programs. This methodology should be applicable to MontiArc models by examining their atomic components, which are already represented by atomic model artifacts. Implementing this approach will necessitate adjustments to the tool architecture.

Additional optimization strategies to consider include \emph{fast unsatisfiability checks}, \emph{common sub-constraint elimination}, and \emph{incremental solving}, as utilized by jCUTE and discussed in \cref{sec:relatedWork}.

Optimization is a critical factor in analyzing realistic models. Our tool is currently in its initial version, and further optimization is essential for practical applications. At present, it supports primitive data types, strings, and enums for MontiArc models. Given that this represents only a fraction of the potential data types, expanding support for additional data types should be prioritized in future work.

A noteworthy observation regarding other DSE tools is that most utilize interpreters for Dynamic Symbolic Execution. However, there is currently no interpreter available for MontiArc models. Developing an interpreter for MontiArc poses several challenges, such as ensuring synchronization and managing the non-sequential execution of all components. For the purposes of this work, we determined that enhancing the MontiArc-to-Java generator would be sufficient. Nevertheless, introducing an interpreter would improve usability, as it would eliminate the need for generating additional files.
    \section{Conclusion} \label{sec:conclusion}
In this paper, we successfully developed a Dynamic Symbolic Execution (DSE) approach tailored for the component-and-connector architecture language MontiArc, implementing multiple execution strategies. This DSE framework enabled us to create a semantic differencing operator capable of detecting behavioral differences between two component-and-connector architectures. Notably, our approach addresses the limitations of previously developed semantic differencing operators by accommodating an infinite state space and input-output alphabet.

We evaluated our DSE approach and the implemented controller based on the criteria of \textit{runtime}, \textit{minimality}, and \textit{completeness}, identifying scalability as the most significant challenge. While our semantic differencing approach is sound, it remains constrained by input length and does not scale well.

Looking ahead, we aim to tackle these challenges using the methods discussed in \cref{sec:discussion}, such as parallelization and strategic early evaluation cessation. We will also reevaluate the updated tool using existing component-and-connector models from both industry and scientific literature.

Furthermore, semantic differencing is not the only application of DSE concerning MontiArc models that interests us. Future work will explore DSE's applicability for test-case and input generation in MontiArc. We are also keen to investigate the highlighting of syntactic differences that lead to semantic differences, expand support for additional data types, and assess the use of an interpreter in comparison to our current generator-based approach. Finally, we may consider applying a DSE-based approach to other types of executable models.

	\begin{acknowledgments}
    Funded by the Deutsche Forschungsgemeinschaft (DFG, German Research Foundation) - 250902306
\end{acknowledgments}

	\bibliography{main.bib}

		
\end{document}